\documentclass[A4paper,9pt,twocolumn,DIV22]{scrartcl}
\usepackage{graphicx}
\usepackage{amssymb}
\usepackage{amsthm}
\usepackage{amsmath}
\usepackage{color}

\newcommand{\figa}{\begin{figure}[!htb]}
\newcommand{\fige}{\end{figure}}

\begin{document}




\title{Pattern formation during the oscillatory photoelectrodissolution of n-type silicon: \\
Turbulence, clusters and chimeras}

\author{Konrad Sch\"onleber, Carla Zensen, Andreas Heinrich and Katharina Krischer}


\maketitle
\begin{abstract}
We report and classify the rich variety of patterns forming spontaneously in the oxide layer during the oscillatory photoelectrodissolution of n-type doped silicon electrodes under limited illumination. Remarkably, these patterns are often comprised of several dynamical states coexisting on the electrode, such as subharmonic phase clusters and spatio-temporal chaos, and include so-called 'chimera states'. The experiments suggest that the subharmonic phase clusters emerge from a period doubling bifurcation which, upon further parameter changes, evolve into classical phase clusters. Experimentally the occurrence of the patterns is controlled via two coupling mechanisms: A linear global coupling by an external resistor and a nonlinear coupling imposed on the system by the limitation of the illumination.  
\end{abstract}






\section{Introduction}

Coupled oscillations are encountered in many contexts in nature and technology. Their prevalence spans over such diverse fields as pacemaker cells in the heart muscle \cite{Cherry:2008}, neural activity in the brain \cite{Izhikevich:2000}, long term climate patterns \cite{Mann:1995} or the electrical power grid \cite{Rohden:2012}. All such systems have in common that their dynamical behavior results from the interaction between a local oscillation and the coupling between the individual oscillators. Thus, the dynamics depends on generic properties of the oscillations such as their harmonical or relaxational character and the way they interact with each other, which might be local (diffusive), long-range or global \cite{Kuramoto:1984}. Therefore, the study of a specific system yields also universal information of spatio-temporal organization of other oscillatory systems \cite{Cross:1993}. In this context, we study an electrochemical oscillatory medium, the photoelectrodissolution of n-doped silicon, with a spatial coupling that produces novel types of synchronization patterns.
\\\\
Our work ties in with previous studies of oscillatory media in chemical and physical systems. Best known are the Belousov-Zhabotinsky (BZ) reaction and the oxidation of CO on Pt in UHV \cite{Zaikin:1970, Epstein:1996, Jakubith:1990, Ertl:1991}, which proved to be prototypical systems with diffusive coupling, exhibiting a variety of dynamically different states, such as sprial waves and chemical turbulence. In addition to the diffusive, i.e. local, coupling, there are many physical mechanisms leading to a long-range or even global coupling of the oscillators. This applies, e.g., to electrically controlled systems, such as semiconductor devices \cite{Meixner:2000, Scholl:2001}, gas discharge tubes \cite{Purwins:2010} or electrochemical systems \cite{Mazouz:1997, Christoph:2002, Krischer:2003, Varela:2005}.
The most prominent patterns emerging from a global coupling are scale free cluster patterns where the medium separates into a few domains, oscillating with a certain, constant phase lag \cite{Falcke:1995, Vanag:2000, Wang:2000, Bertram:2003, Varela:20052}.
\\\\
In parallel to these experimental works, theoretical studies employing the complex Ginzburg-Landau equation (CGLE) or phase oscillator models as universal descriptions of oscillatory systems were carried out \cite{Pikovsky:2001, Aranson:2002}. The former is a normal form equation of an oscillatory medium close to the Hopf bifurcation where the limit cycles are only weakly attractive, while the latter captures the dynamics of strongly attractive, coupled limit cycles. Among the vast literature on the solutions of these equations, cluster patterns are most relevant in the context of our work \cite{Hakim:1992, Okuda:1993}. Typically, one distinguishes between phase clusters, where all oscillators follow the same limit cycle but form distinct groups with a constant phase difference, and amplitude clusters, where the different groups occupy different limit cycles with distinct amplitudes and phases. Both types of clusters emerge in globally coupled systems, however, phase clusters require coupling functions with higher harmonics. Thus, in contrast to amplitude clusters, they cannot be described in the framework of the CGLE with a linear global coupling. In our work we encounter cluster patterns with novel features, most remarkably self-organized chimera states and other types of coexistence of different dynamical states on the electrode surface, and we discuss that a nonlinear coupling leads to their emergence. Furthermore, a collection of other peculiar patterns and unusual synchronization phenomena is presented.
\\\\
Our experimental system is the photoelectrochemical oxidation of n-doped silicon to SiO$_2$ under potentiostatic conditions and the concurrent dissolution of this oxide mediated by fluoride species. Owing to this interplay of oxidation and etching process the system exhibits oscillations in current and oxide layer thickness in a broad parameter regime \cite{Turner:1958, Gerischer:1988, Zhang:2001}.  
The electrochemical oxidation equation reads:
\begin{equation}
\mathrm{Si}+2\mathrm{H_2O}+\nu_{\mathrm{VB}}h^+\rightarrow \mathrm{SiO}_2+4\mathrm{H^+}+(4-\nu_{\mathrm{VB}})e^-
\label{eq:OxReac}
\end{equation}
where $\nu_{\mathrm{VB}}$ denotes the number of charge transfer processes via the valence band of the silicon, i.e., processes involving minority charge carriers in n-doped silicon. These holes drive the initial charge transfer step in equation \ref{eq:OxReac}, i.e., the inequality $\nu_{\mathrm{VB}}\geq 1$ holds and there is no oxidation current without illumination \cite{Hasegawa:1988}. A limited illumination and, therefore, limited generation rate of holes in the valence band of the working electrode thus introduces a cut-off for the total current. It constitutes a global constraint as well as a nonlinear coupling mechanism between the oscillating electrode positions.\\
An additional, global coupling mechanism is introduced by connecting a resistance $R_{\mathrm{ext}}$ in series to the working electrode in the potentiostatically controlled experiments. This external resistance couples the voltage drop $\Delta\phi_{\mathrm{int}}$ across the electrode$|$oxide$|$electrolyte interface at each individual point of the surface of the working electrode to the total current through the working electrode and the introduced coupling consequently has a linear characteristic.
\begin{equation}
\Delta\phi_{\mathrm{int}}=U-R_{\mathrm{ext}}A\overline{j}
\label{eq:linglobcoup}
\end{equation}
Here, $U$ is the externally applied voltage, $\overline{j}$ the spatial average of the current density and $A$ the electrode area.\\\\ 
In order to get a comprehensive picture of the dependence of the patterns on the experimental parameters, four of these parameters are varied over a wide range where current oscillations occur. Two of these parameters characterize the electrolyte, the first one being the total concentration of fluoride species $c_{\mathrm{F}}$ and the second one the pH value which determines the distribution of the fluoride on the three species F$^-$, HF and HF$_2^-$ according to the law of mass action. Both parameters together determine the total silicon oxide etching rate as the different fluoride species follow different etching pathways with distinct rates \cite{Judge:1971, Cattarin:2000}. The other two parameters represent the strengths of the coupling mechanisms given by the illumination intensity $I_{\mathrm{ill}}$ for the nonlinear coupling and $R_{\mathrm{ext}}A$ for the purely linear global coupling, respectively.\\\\ 
After a description of the experimental setup for the spatially resolved recording of the oxide layer with an ellipsometric imaging technique and the data processing in section \ref{sec:Exp}, an overview of the parameter dependence of the pattern forming regime is given and the patterns occurring in the experiments are classified by spatially averaged quantities. Subsequently, in sections \ref{subsec:P1}-\ref{subsec:Irreg} for each such class the different prototypic patterns encountered in the experiments are shown in a systematic way. Additionally, the requirements on a generic model exhibiting such spatio-temporal dynamics are discussed.
\section{Experimental}
\label{sec:Exp}
\begin{figure}[ht]
	\begin{center}
		\includegraphics[width=0.35\textwidth]{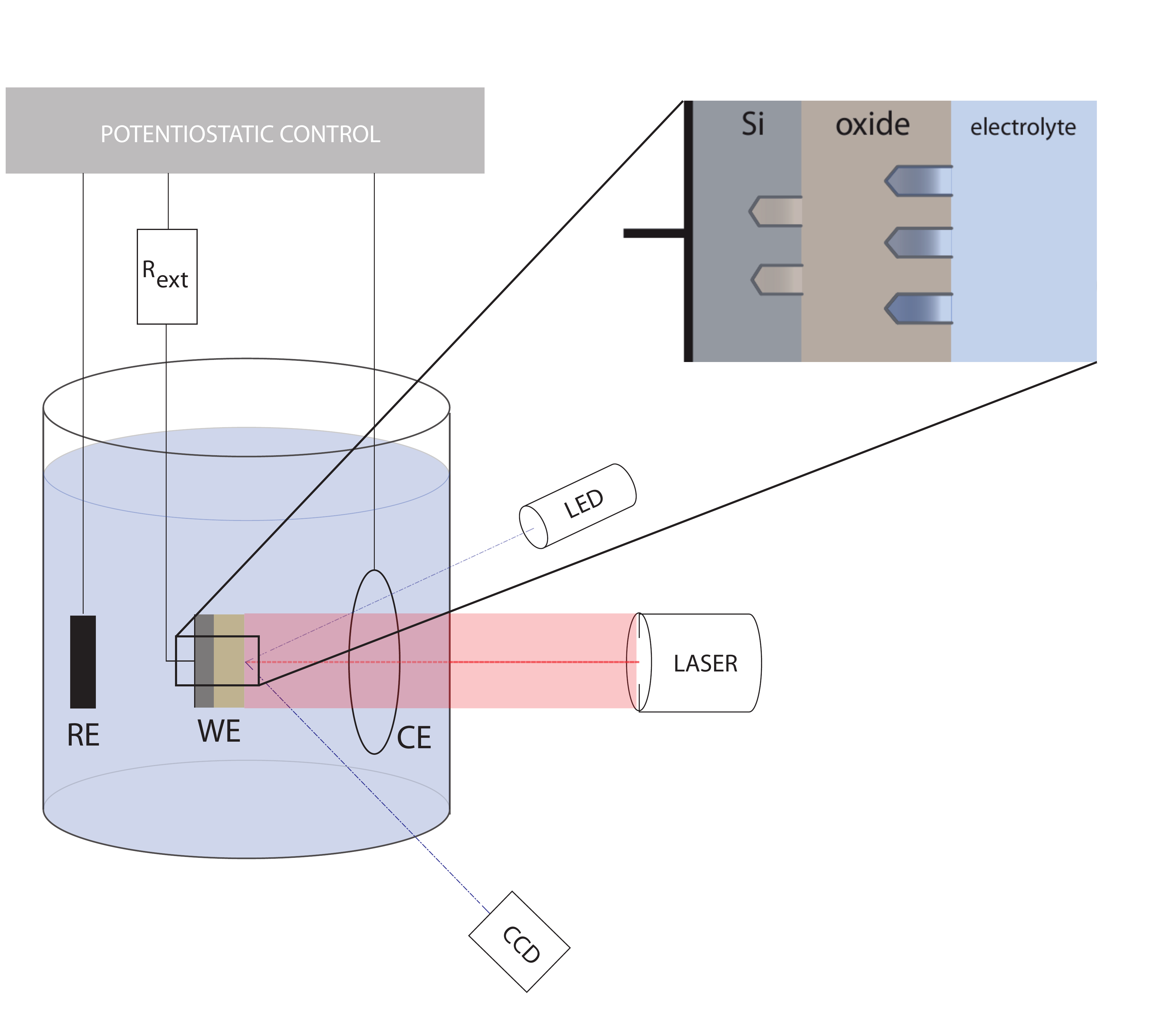}
	\end{center}
		\caption{Sketch of the experimental setup showing the arrangement of the three electrodes and the external resistance together with the optical paths for the sample illumination (red) and the spatially resolved ellipsometric imaging (blue). A cross section of the interface and the growth direction of the oxide is shown in the inset. A schematic of the optical components in the light path between the LED and the CCD can be found in \cite{Miethe:2012}}
	\label{fig:Sketch}
\end{figure}

The working electrodes are the (111) faces of monocrystalline, n-type doped silicon samples with a resistivity of $3-5~\Omega$cm. They are ohmically back-contacted with a thermally evaporated 200 nm aluminum layer annealed at $250~^{\circ}$C for $15$ minutes. The silicon is attached to a custom made PTFE holder via silicone paste (Scrintex 901, Ralicks GmbH, Rees-Haldern, Germany) which also serves as a sealing for the back-contact against the electrolyte. The typical electrode size is $A\approx4\times6$ mm$^2$. As a counter electrode, a circular shaped platinum wire is placed symmetrically in front of the working electrode and a saturated Hg$|$Hg$_2$SO$_4$ reference electrode is used which is placed behind the working electrode at a distance of several centimeters. During the experiments the electrolyte is constantly stirred with a magnetic stirrer rotating at about 10 Hz. The sample illumination is realized by a He-Ne laser and the illumination intensity is adjusted by a polarizer plate. The laser beam is optically expanded to a diameter of ca. 1.5 cm and passes an iris diaphragm so that only its central part is incident on the sample to guarantee a spatially uniform illumination.\\\\
For the ellipsometric imaging of the optical path through the oxide, light from a blue LED ($\lambda=470$ nm) is first elliptically polarized by passing a linear polarizer and a zeroth order $\lambda$/4-plate. It then hits the sample surface at an angle close to the Brewster angle of the silicon$|$electrolyte system ($\approx70^{\circ}$). The reflected light passes another polarizer which converts changes in the polarization upon reflection caused by the optical path through the oxide into an intensity signal. Using a suitable lens, an image of this ellipsometric intensity distribution across the surface is then created on the CCD chip of a camera (JAI CV-A50). The data is then digitized using a frame grabber card (PCI-1405, National Instruments) and preprocessed and recorded by a Labview program. A sketch of the setup (with the optical components of the ellipsometric imaging path omitted) is shown in figure \ref{fig:Sketch}.\\\\
Before the experiments the silicon working electrodes are cleaned by first rubbing them carefully with an acetone soaked tissue and subsequently immersing them in acetone (p.a.), ethanol (p.a.), methanol (p.a.) and ultrapure water for 5 min each. The electrolyte is always an aqueous solution of NH$_4$F (Merck, p.a.) as a fluoride source and H$_2$SO$_4$ (Merck, Suprapur) to adjust the pH value. Prior to the experiments the electrolyte is bubbled with argon and an argon atmosphere is retained during the experiments.\\\\
All experiments are potentiostatically controlled (Potentiostat: FHI-2740, electronics laboratory of the Fritz-Haber-
Institut, Berlin, Germany). A constant voltage of 8.65 V vs. standard hydrogen electrode is applied between the reference and the working electrode with a tunable ohmic resistance connected in series to the working electrode.       
\\\\
For the analysis of the patterns the spatial distribution of the ellipsometric intensity at specified points in time is considered as well as the ellipsometric signal of one dimensional subsets of the electrode over the entire time window. Furthermore, the time series of the ellipsometric intensity at each individual point is Fourier transformed (FFT, MATLAB) and the spatial distribution of absolute value and phase of relevant Fourier coefficients is then used to gain insight into phase clustering behavior for these Fourier modes. Lastly, the data from a selected area on the electrode is Hilbert transformed (MATLAB) and the resulting analytical signal, i.e. the complex signal with the original data as the real part and the Hilbert transform of the data as the imaginary part, is used to visualize the different dynamics on the electrode.
\section{Results and Discussion}
\subsection{Overview}
\subsubsection*{Parameter variation:}
\begin{figure}[ht]
	\centering
		\includegraphics[width=0.35\textwidth]{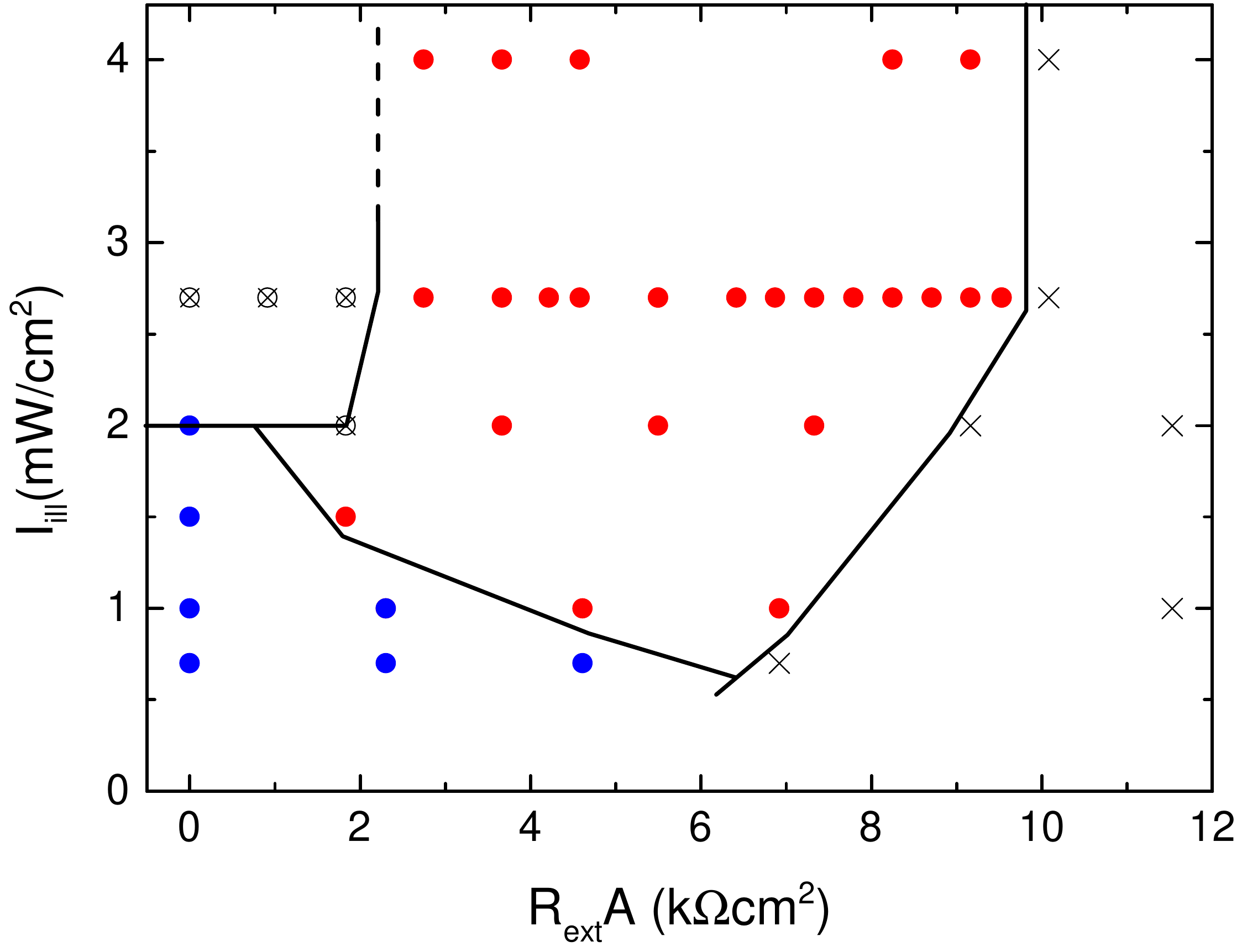}
		\caption{Dependence of the oscillatory and pattern forming regime on the external resistance $R_{\mathrm{ext}}A$ and the illumination intensity $I_{\mathrm{ill}}$ for an electrolyte with $c_{\mathrm{F}}$=50 mM and pH 2.3. Uniform oscillations (red) and oscillations accompanied by spatial pattern formation (blue) are shown together with stable fixed points with (circle) and without (cross) an oscillatory transient outside of the oscillatory regime. The borders of the oscillatory and the pattern forming regime are indicated by lines.}
	\label{fig:Parameter}
\end{figure}

\begin{figure}[ht]
	\centering
		\includegraphics[width=0.35\textwidth]{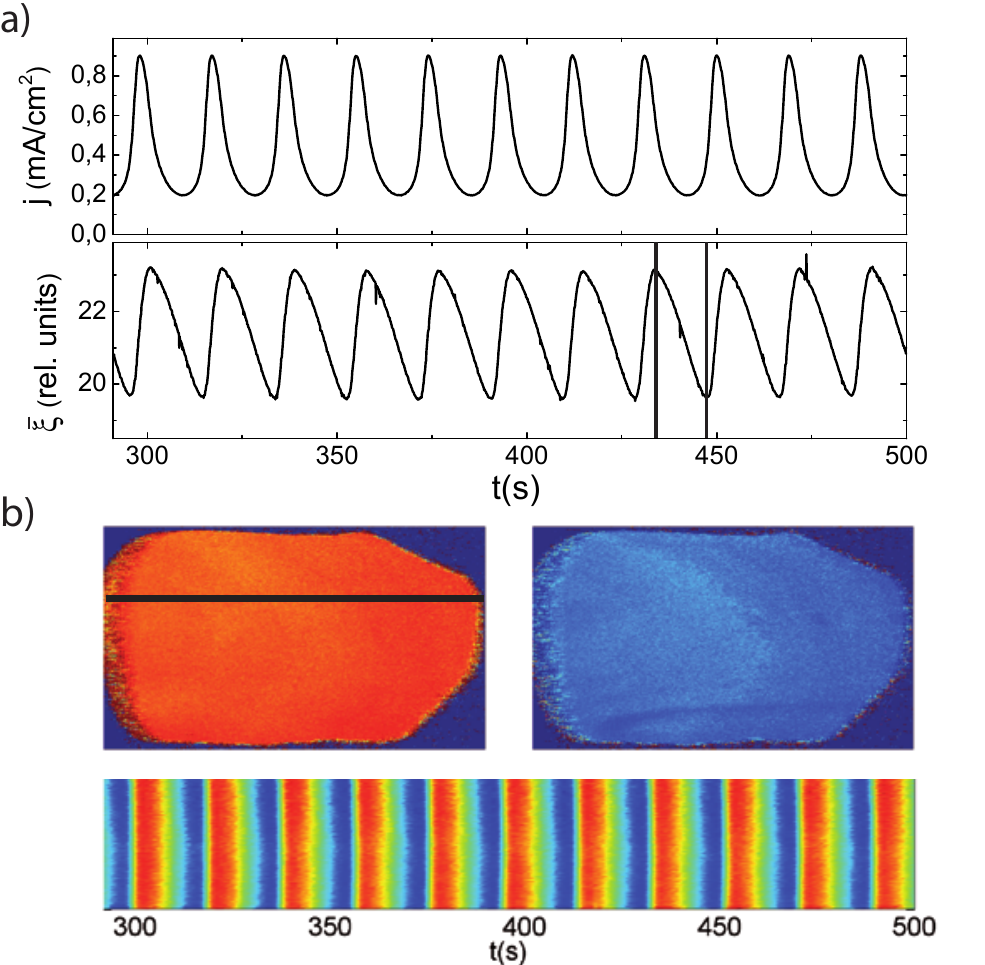}
		\caption{Uniform oscillation of the electrode at a high illumination intensity ($c_{\mathrm{F}}=75$ mM, $\mathrm{pH}=1$, $R_{\mathrm{ext}}A=6.51~\mathrm{k\Omega cm^2}$, $I_{\mathrm{ill}}=5.85$ mW/cm$^2$). \textbf{a)} Time series of the global quantities $j$ and $\overline{\xi}$; \textbf{b)} Ellipsometric intensity distribution on the electrode for the two times indicated by the vertical lines in a) and the temporal evolution of a 1d cut along the line indicated in the left electrode picture. Red indicates a relatively high and blue a relatively low ellipsometric intensity.}
	\label{fig:NurHom1}
\end{figure}

The occurrence of oscillations and the accompanying pattern formation for a given electrolyte depends on the values of the scaled external resistance $R_{\mathrm{ext}}A$ and the illumination intensity $I_{\mathrm{ill}}$. An exemplary 2d parameter space is shown in figure \ref{fig:Parameter}.
At high $I_{\mathrm{ill}}$ and low $R_{\mathrm{ext}}A$, the system settles to a stable focus. Conversely, if the external resistance is set at a too high value, the system relaxes monotonically to a stable node. In between, within the oscillatory regime, both parameters have an impact on the local oscillators, but as explained above they mediate also a coupling between the individual positions on the electrode. The impact on local dynamics and spatial coupling cannot be separated in the experiments. However, as we elaborate below, we obtain a consistent picture if we assume that the principle trend of the respective locations of patterns and synchronized oscillations in the parameter plane is dominated by the two spatial coupling mechanisms rather than the change in the properties of the local oscillators.\\\\    
For relatively high illumination intensities, i.e at a low nonlinear coupling strength, the entire electrode oscillates uniformly reflecting a strong synchronizing spatial coupling. An example for such a uniform oscillation is shown in figure \ref{fig:NurHom1}.
Uniform oscillations are also found in all oscillatory measurements at non-illuminated p-type silicon samples,  as discussed in detailed studies of such samples conducted previously in our group \cite{Miethe:2012, Schoenleber:2012}. In addition to the uniformity, the shape of the oscillations for highly illuminated n-type doped silicon and p-type doped silicon samples are identical at otherwise matched experimental parameters and in both cases a minimal external resistance has to be introduced. This similarity in behavior can be fully attributed to the sufficiency of the amount of holes in the valence band to maintain the current present in both cases, strongly corroborating the notion that the nonlinear coupling introduced by the restriction of holes is solely responsible for the emergence of patterns. Consequently, in the parameter space shown in figure \ref{fig:Parameter} only at illumination intensities below an electrolyte specific threshold value of $I_{\mathrm{ill}}=2.5$ mW/cm$^2$ pattern formation can be observed. If the linear global coupling is too strong, the patterns vanish and uniform oscillations are observed although the current is still restricted by the illumination and an effect of the nonlinear coupling is thus still to be expected. This behavior is in accordance with literature where the global coupling through an external resistor tends to synchronize the oscillations, at least when the electrode potential acts as the activator \cite{Krischer:2000,Krischer:2002}. As a consequence patterns are only found in the parameter region where the nonlinear coupling dominates the linear global coupling, i.e., in the lower left hand corner of figure \ref{fig:Parameter}. Most notably, here, sustained oscillations are observed even without an external resistance.
\subsubsection*{Pattern classification:}
\begin{figure}[ht]
	\centering
		\includegraphics[width=0.35\textwidth]{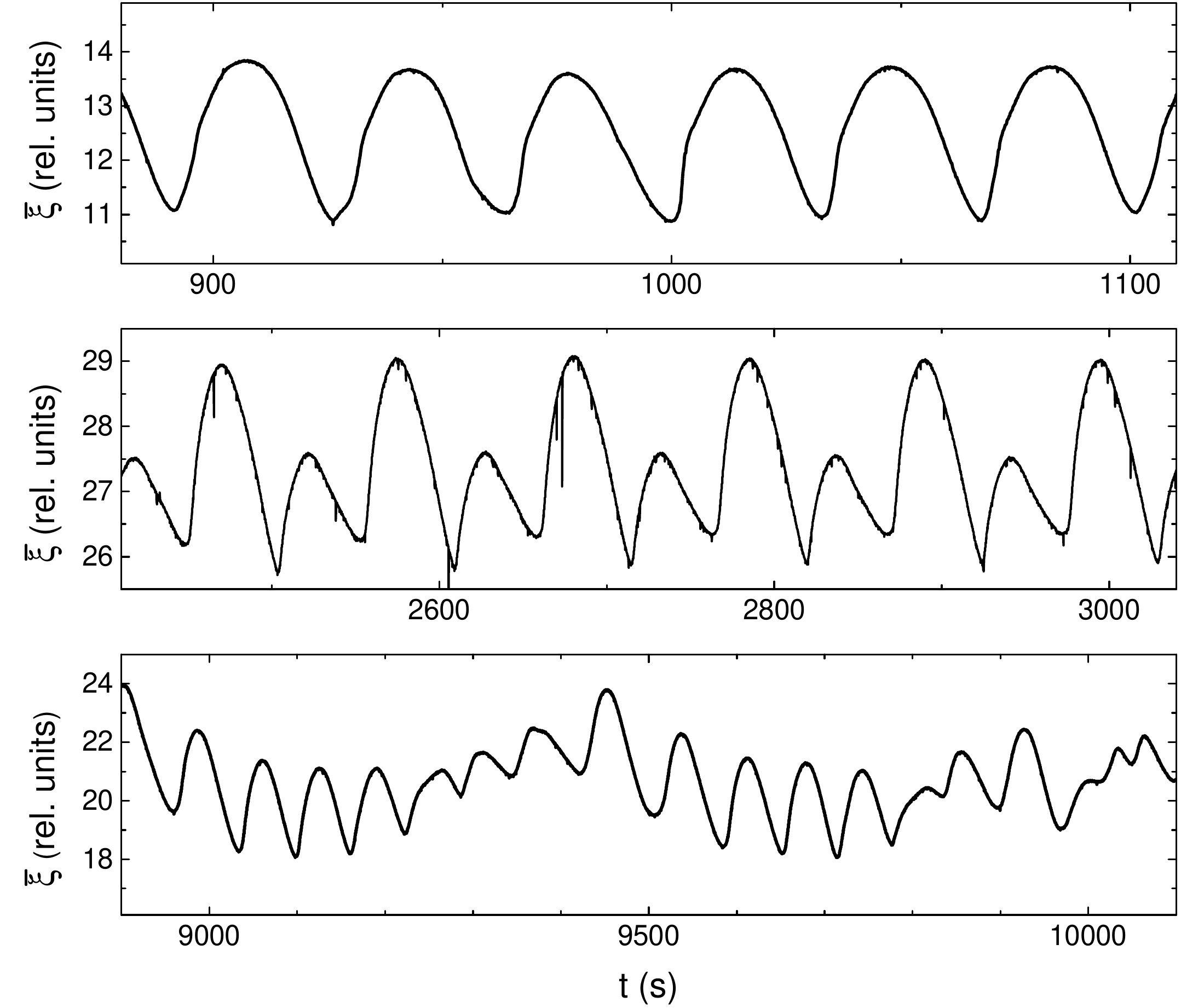}
		\caption{Time series of the spatially averaged ellipsometric intensity $\overline{\xi}$ for three different measurements showing simple periodic oscillations (top), period-2 oscillations (middle) and oscillations with an irregular amplitude (bottom).}
	\label{fig:Classification}
\end{figure}

Compared to patterns classically found in oscillatory media the silicon oscillator exhibits two characteristic and peculiar phenomena: First, in a large part of the pattern forming parameter region the global quantities, i.e., the total current $j$ and the spatially averaged ellipsometric intensity $\overline{\xi}$ oscillate in a simple periodic fashion although local time series exhibit complex oscillation forms up to completely irregular ones. Besides, $\overline{\xi}$ may oscillate with a period-2 or irregularly. Second, typically different dynamic behaviors as for example cluster patterns and spatio-temporal chaos coexist on the electrode. We call a part of the surface with one distinct dynamics a region.\\\\ 
In the following we classify the patterns, in a first step, according to the spatially averaged ellipsometric intensity $\overline{\xi}$ into three categories:
\begin{itemize}
\item Simple periodic oscillations
\item Period-2 oscillations
\item Oscillations with an irregular amplitude
\end{itemize} 
Exemplary time series for the three types of $\overline{\xi}$ oscillations are shown in figure \ref{fig:Classification}. 
In the case of simple periodic oscillations in $\overline{\xi}$, 2d simulations using the suitably modified complex Ginzburg-Landau equation with a nonlinear global coupling show striking similarities with the patterns observed in the experiments \cite{Miethe:2009, Garcia-Morales:2010, Schmidt:2014}. This opens a pathway to a possible deeper understanding of both the origin of the pattern formation and the coupling mechanism. Consequently, in the following the emphasis is first put on the analysis of the first class of patterns and the second and third class are treated as modifications of this first class and are discussed later.\\\\
           
\subsection{Simple periodic oscillations in $\overline{\xi}$}
\label{subsec:P1}
In a first step we now consider the patterns forming during simple periodic $\overline{\xi}$ oscillations depicted in the top of figure \ref{fig:Classification}. The regularity of the spatially averaged signal implies that the spatial variations of the patterns cancel out on average. There are two dominant types of patterns forming. First, subharmonic phase clusters are found, where the ellipsometric intensity at each point shows oscillations with a strong frequency component at an integer fraction of the spatially averaged oscillation frequency. These points are arranged in phase domains where all points within a domain are synchronized and the distinct domains differ in their phase \cite{Miethe:2009}. Typically but not exclusively, 2-phase clusters are found.
The second type of dynamical behavior found are desynchronized oscillations with a broad distribution of oscillation frequencies. Hence, at any given point the oscillation amplitude is irregular.\\\\
As already mentioned, typically the electrode forms different regions. In a region either one of the two patterns mentioned or a synchronized oscillation with the frequency of the spatially averaged signal exists. The occurrence of only one type of dynamical behavior on the entire electrode surface and any pairwise coexistence have been found in this study and are subsequently presented.
\subsubsection*{One region}
\begin{figure}[ht]
	\centering
		\includegraphics[width=0.35\textwidth]{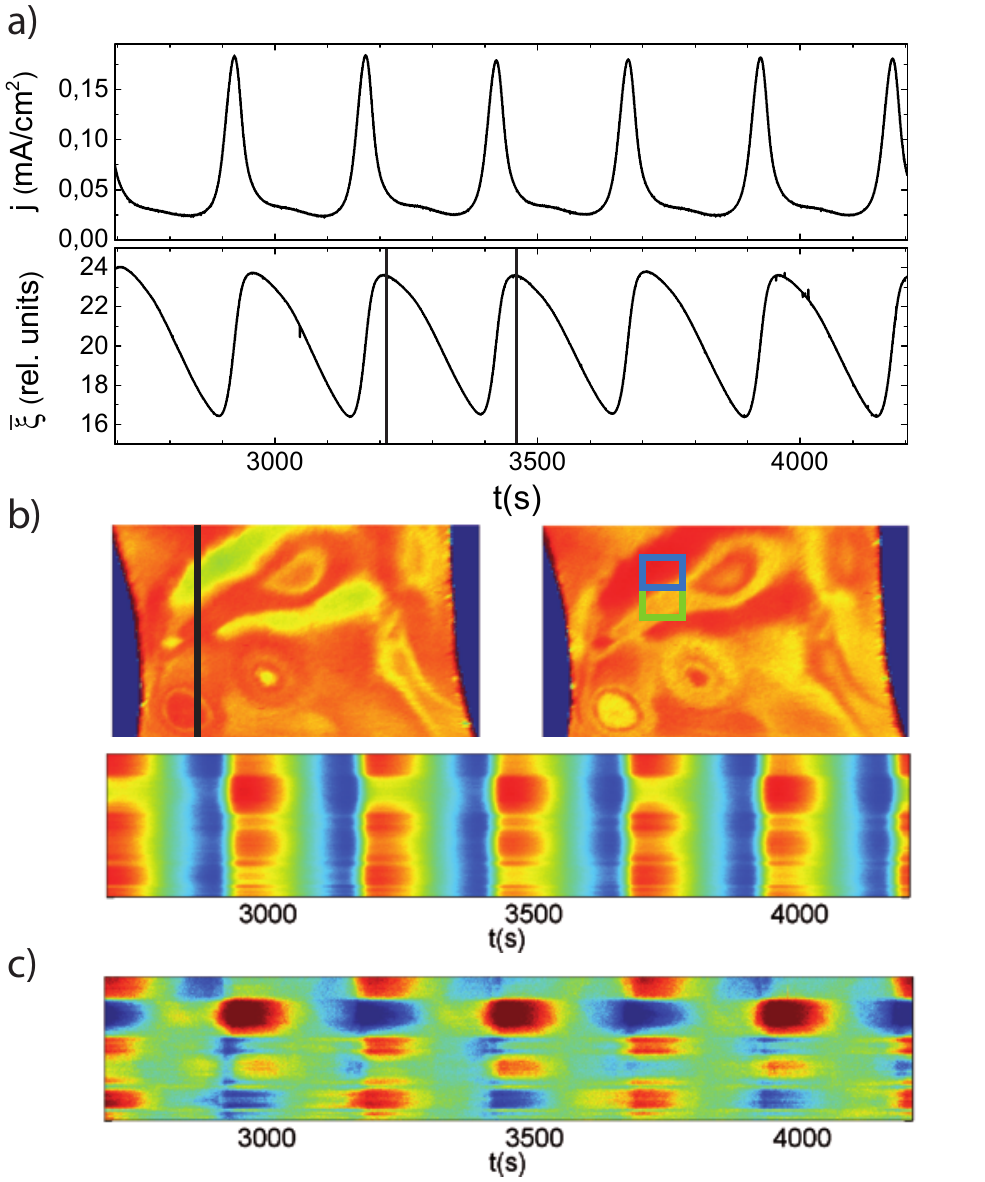}
		\caption{Subharmonic cluster pattern covering the entire electrode ($c_{\mathrm{F}}=35$ mM, $\mathrm{pH}=1$, $R_{\mathrm{ext}}A=9.1~\mathrm{k\Omega cm^2}$, $I_{\mathrm{ill}}=0.7$ mW/cm$^2$). \textbf{a)} Time series of $j$ and $\overline{\xi}$; \textbf{b)} Ellipsometric intensity distribution on the electrode for the two times indicated by the vertical lines in a) and the temporal evolution of a 1d cut along the line indicated in the left electrode picture. Red indicates a relatively high and blue a relatively low ellipsometric intensity; \textbf{c)} 1d cut as in b) where $\overline{\xi}$ is subtracted from every local time series. The color scale is not identical to b).}
	\label{fig:NurSH1}
\end{figure}

\begin{figure}[ht]
	\centering
		\includegraphics[width=0.35\textwidth]{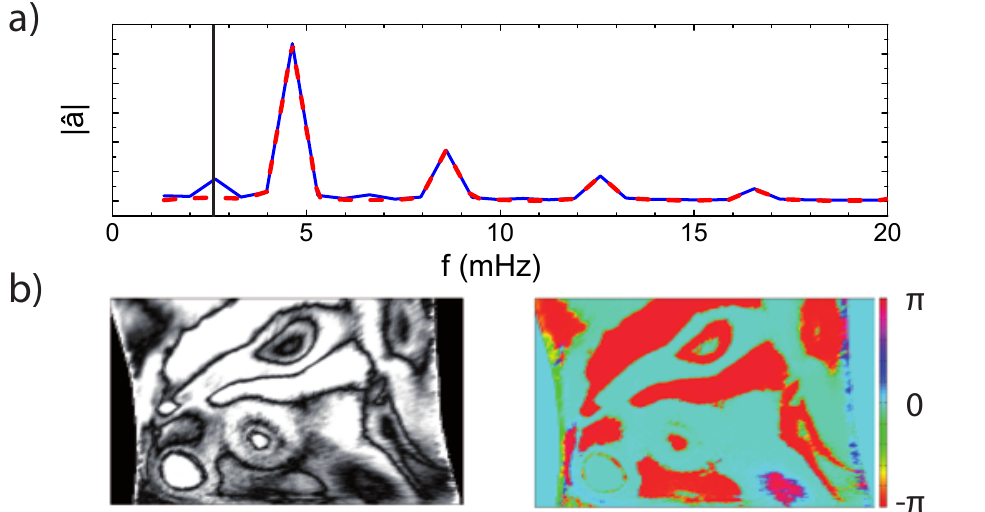}
		\caption{Fourier analysis of the subharmonic cluster pattern shown in figure \ref{fig:NurSH1}. \textbf{a)} Spatial average of the absolute values of the pointwise Fourier spectra (blue) and Fourier spectrum of the spatially averaged signal $\overline{\xi}$ (red); \textbf{b)} Spatial distribution of the absolute value (left) and phase (right) of the Fourier coefficient corresponding to the Fourier mode indicated by the vertical line in the Fourier spectra in a).}
	\label{fig:NurSH2}
\end{figure}

\begin{figure}[ht]
	\centering
		\includegraphics[width=0.35\textwidth]{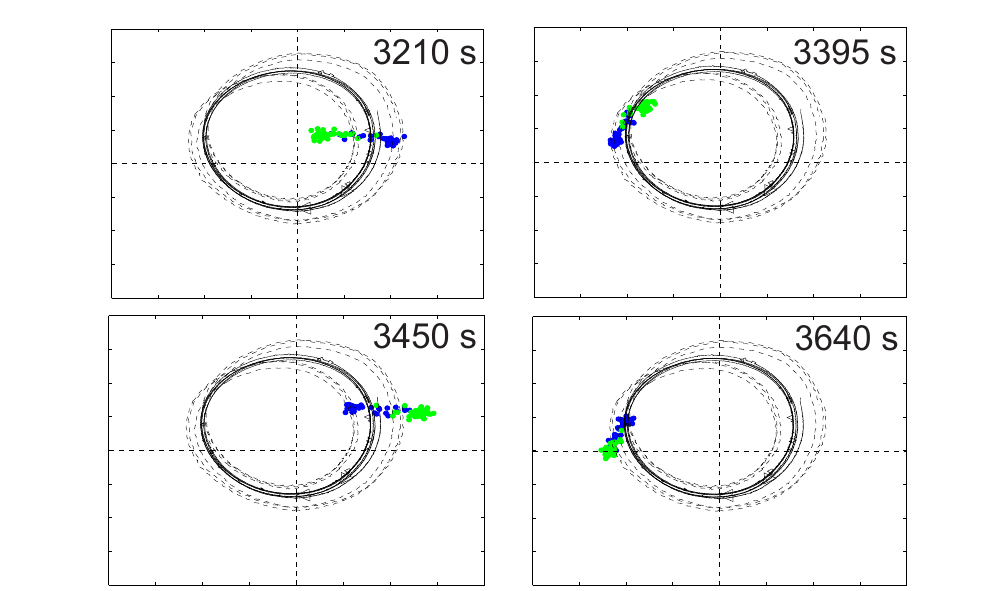}
		\caption{Analytical signal from the points indicated by correspondingly color coded rectangles in figure \ref{fig:NurSH1} b) at four consecutive extrema of the $\overline{\xi}$ oscillation shown in figure \ref{fig:NurSH1} a) together with the traces of the analytical signal of $\overline{\xi}$ (solid) and a point located in the center of the blue rectangle in figure \ref{fig:NurSH1} (dash).}
	\label{fig:NurSH3}
\end{figure}

\begin{figure}[ht]
	\centering
		\includegraphics[width=0.35\textwidth]{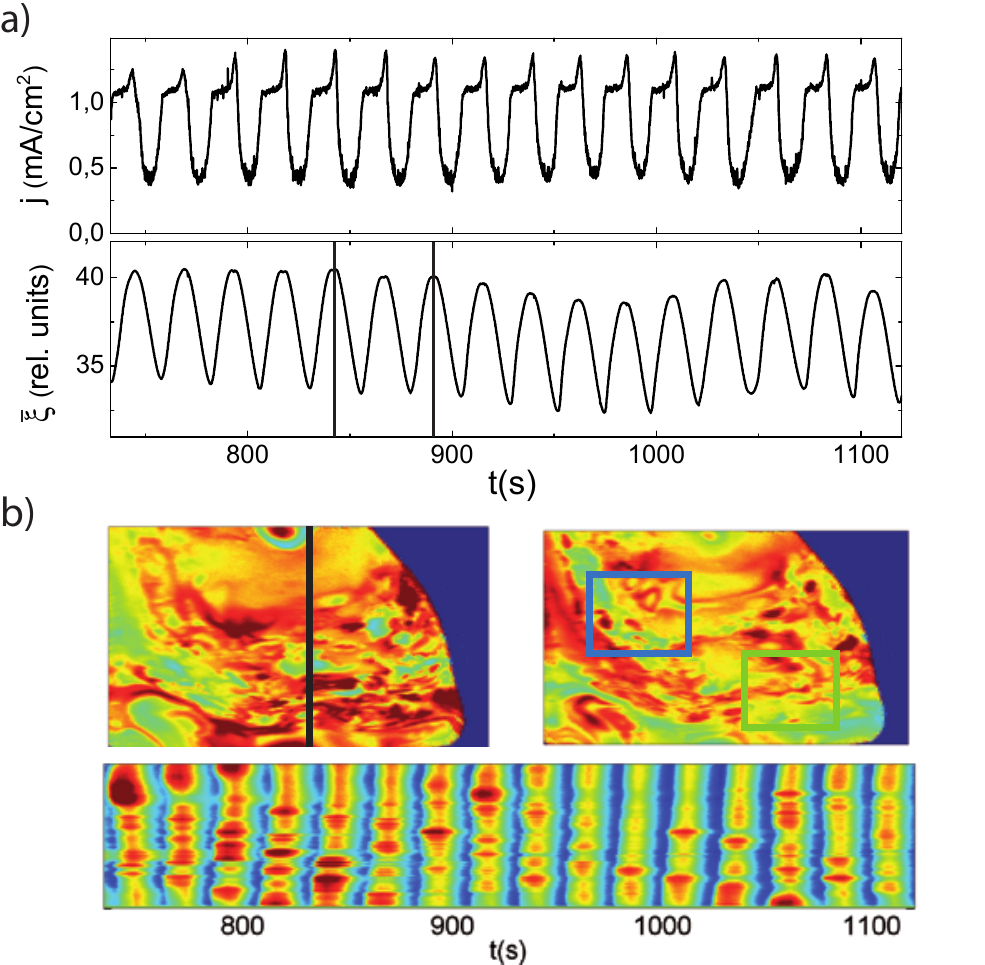}
		\caption{Desynchronized oscillation pattern covering the entire electrode ($c_{\mathrm{F}}=75$ mM, $\mathrm{pH}=1$, $R_{\mathrm{ext}}A=0~\mathrm{k\Omega cm^2}$, $I_{\mathrm{ill}}=2.0$ mW/cm$^2$). \textbf{a)} Time series of $j$ and $\overline{\xi}$; \textbf{b)} Ellipsometric intensity distribution on the electrode for the two times indicated by the vertical lines in a) and the temporal evolution of a 1d cut along the line indicated in the left electrode picture. Red indicates a relatively high and blue a relatively low ellipsometric intensity.}
	\label{fig:NurTurb1}
\end{figure}

\begin{figure}[ht]
	\centering
		\includegraphics[width=0.35\textwidth]{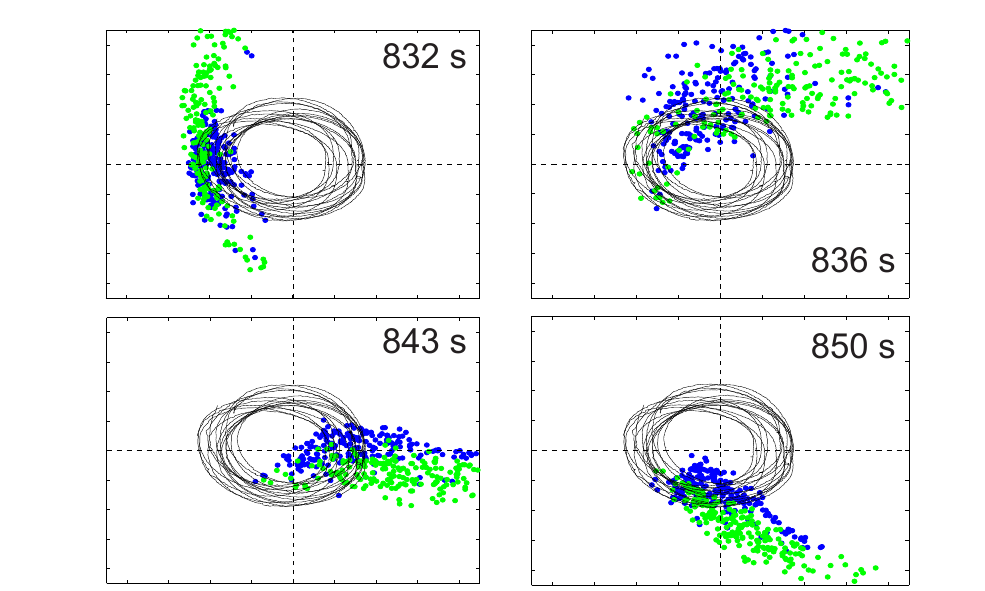}
		\caption{Analytical signal from the points indicated by correspondingly color coded rectangles in figure \ref{fig:NurTurb1} b) at four points during one oscillation period of the $\overline{\xi}$ oscillation shown in figure \ref{fig:NurTurb1} a) together with the trace of the analytical signal of $\overline{\xi}$ (solid).}
	\label{fig:NurTurb2}
\end{figure}

First, we take a look at the patterns with only one region covering the entire electrode. We start by discussing 2-phase clusters. An example is shown in figure \ref{fig:NurSH1}.
As apparent from figure \ref{fig:NurSH1} a) the spatially averaged ellipsometric intensity $\overline{\xi}$ as well as the spatially averaged current density $j$ oscillate in a simple periodic fashion. At the same time this oscillation is not uniform as can be seen in figure \ref{fig:NurSH1} b) where two snapshots of the ellipsometric images and the temporal evolution of a 1d cross section are depicted. Clearly, at every point, an amplitude variation can be seen which inverts every other oscillation. For a better visibility of the spatial distribution of the amplitude variations the underlying uniform oscillation is subtracted in subfigure \ref{fig:NurSH1} c), uncovering subharmonic local oscillations. The characteristics of this structure are further elucidated by the results of a Fourier analysis of the local timeseries at each point of the surface shown in figure \ref{fig:NurSH2}.
In subfigure \ref{fig:NurSH2} a), the spatial average of the absolute values of the pointwise Fourier spectra is compared to the Fourier spectrum of the spatially averaged ellipsometric intensity. The occurrence of an additional peak at half the dominant frequency can clearly be seen. The spatial distribution of absolute value and phase of the Fourier coefficient corresponding to this additional, subharmonic frequency are shown in figure \ref{fig:NurSH2} b). The latter is composed of two values which are about $\pi$ apart, confirming that a 2-phase cluster spreading across the entire electrode area is indeed present. The two clustered phase domains are separated by Ising-type walls, i.e., by a zero crossing of the absolute value of the Fourier coefficient where the phase jump occurs. This manifests in the black lines in the left hand picture of figure \ref{fig:NurSH2} b).\\\\
More insight into the details of the cluster dynamics can be gained by considering the analytical signal of points from the individual phase domains as shown in figure \ref{fig:NurSH3}.
The analytical signal of the points in the two areas indicated in figure \ref{fig:NurSH1} are arranged on a bar (blue and green points, respectively) whose center follows the analytical signal of $\overline{\xi}$ (black solid line). The end points of the bar, i.e., the points from the individual phase domains, trace a Moebius strip projected on the complex plane. This suggests that the subharmonic clusters result from a period doubling bifurcation of the uniform oscillation. The four snapshots in figure \ref{fig:NurSH3} are taken at consecutive extrema of $\overline{\xi}$ and the sequence of pictures thus shows that a full rotation of the bar is accompanied by two full rotations of its center.\\\\
The second type of dynamical pattern covering the entire active area observed is a desynchronized oscillation spreading across the entire electrode as shown in figure \ref{fig:NurTurb1}.     
Here, the remarkable case is shown where the ensemble of pointwise time series of the ellipsometric intensity across the surface is oscillating in a desynchronized manner, while in the spatially averaged signal these differences cancel out almost perfectly leaving a simple periodic oscillation in $\overline{\xi}$ and $j$. The corresponding full analytical signal of two sets of points on the electrode further illustrates this behavior. It is shown in figure \ref{fig:NurTurb2}.
The trajectories of the individual points are uncorrelated, both within the two areas shown and between them, leading to broad point clouds circling the origin.
\subsubsection*{Two regions}
\begin{figure}[ht]
	\centering
		\includegraphics[width=0.35\textwidth]{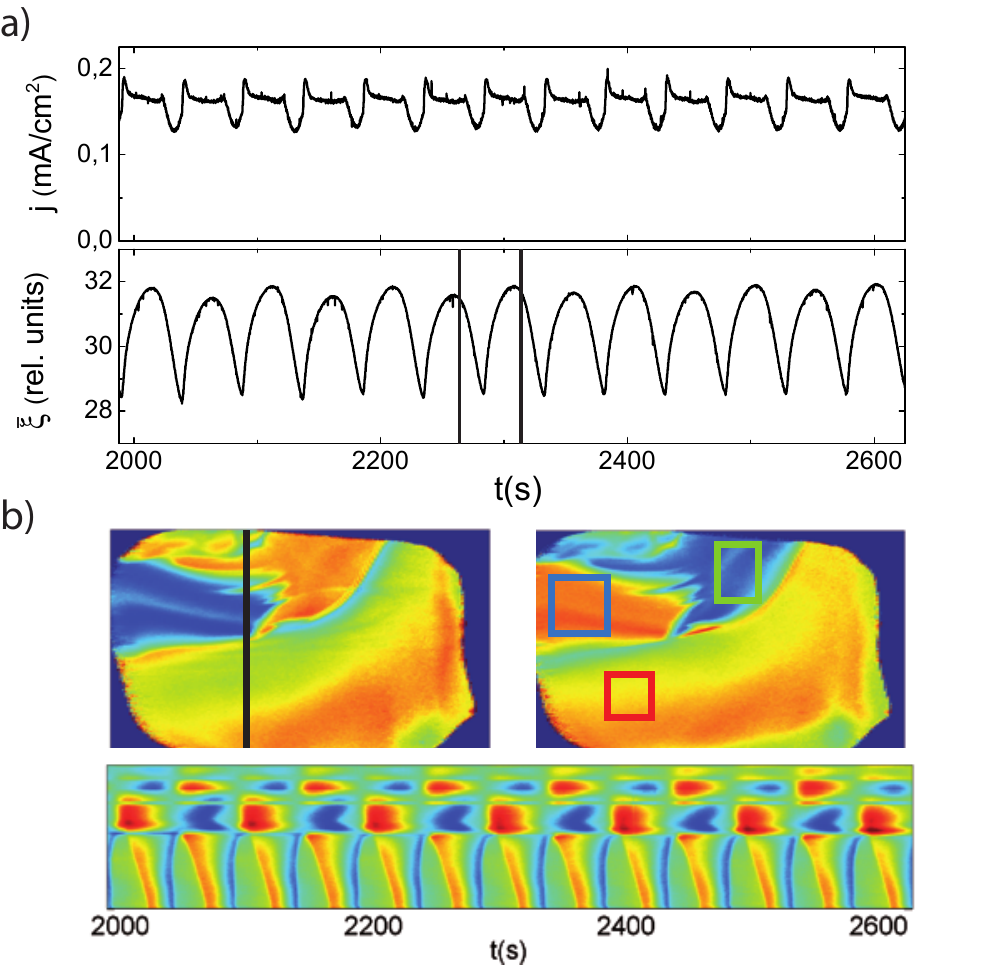}
		\caption{Analysis of a pattern consisting of a synchronized region coexisting with a subharmonic 2-phase cluster ($c_{\mathrm{F}}=50$ mM, $\mathrm{pH}=1$, $R_{\mathrm{ext}}A=0.69~\mathrm{k\Omega cm^2}$, $I_{\mathrm{ill}}=0.4$ mW/cm$^2$). \textbf{a)} Time series of $j$ and $\overline{\xi}$; \textbf{b)} Ellipsometric intensity distribution on the electrode for the two times indicated by the vertical lines in a) and the temporal evolution of a 1d cut along the line indicated in the left electrode picture. Red indicates a relatively high and blue a relatively low ellipsometric intensity.}
	\label{fig:Hom+SH1}
\end{figure}

\begin{figure}[ht]
	\centering
		\includegraphics[width=0.35\textwidth]{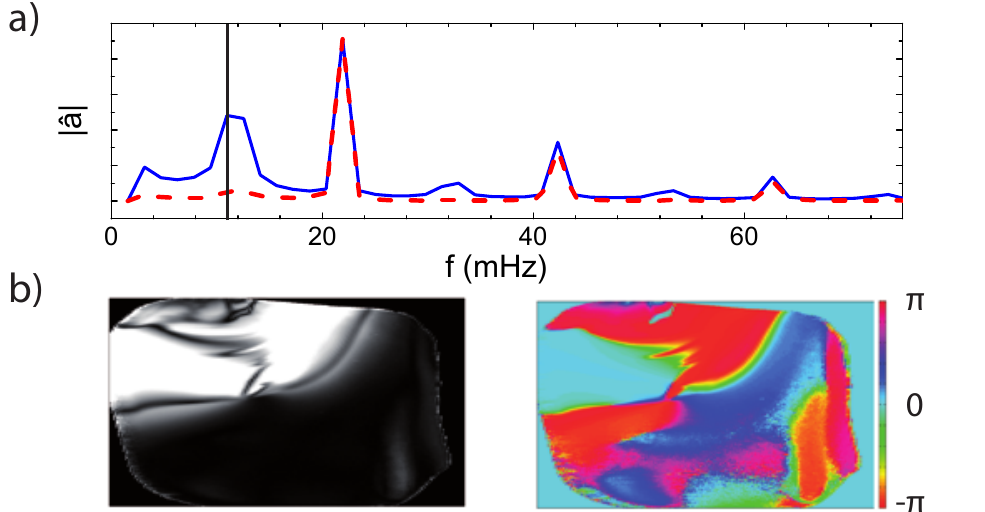}
		\caption{Fourier analysis of the pattern shown in figure \ref{fig:Hom+SH1}. \textbf{a)} Spatial average of the absolute values of the pointwise Fourier spectra (blue) and Fourier spectrum of the spatially averaged signal $\overline{\xi}$ (red); \textbf{b)} Spatial distribution of the absolute value (left) and phase (right) of the Fourier coefficient corresponding to the Fourier mode indicated by the vertical line in the Fourier spectra in a).}
	\label{fig:Hom+SH2}
\end{figure}

\begin{figure}[ht]
	\centering
		\includegraphics[width=0.35\textwidth]{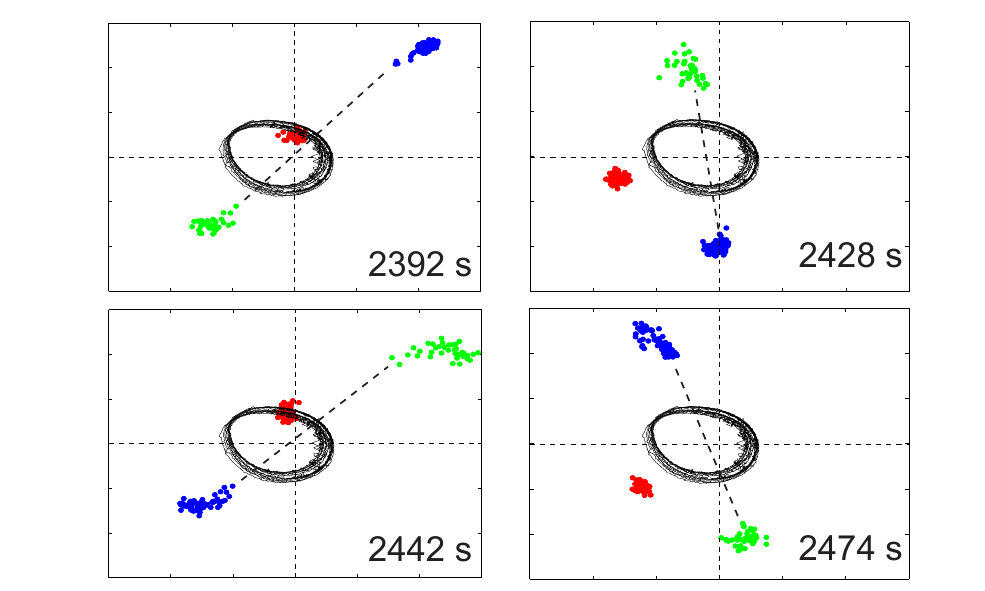}
		\caption{Analytical signal from the points indicated by correspondingly color coded rectangles in figure \ref{fig:Hom+SH1} b) at four points during two oscillation periods of the $\overline{\xi}$ oscillation shown in figure \ref{fig:Hom+SH1} a) together with the trace of the analytical signal of $\overline{\xi}$ (solid). The dashed line represents points in the Ising wall between the two phase domains.}
	\label{fig:Hom+SH3}
\end{figure}

\begin{figure}[ht]
	\centering
		\includegraphics[width=0.35\textwidth]{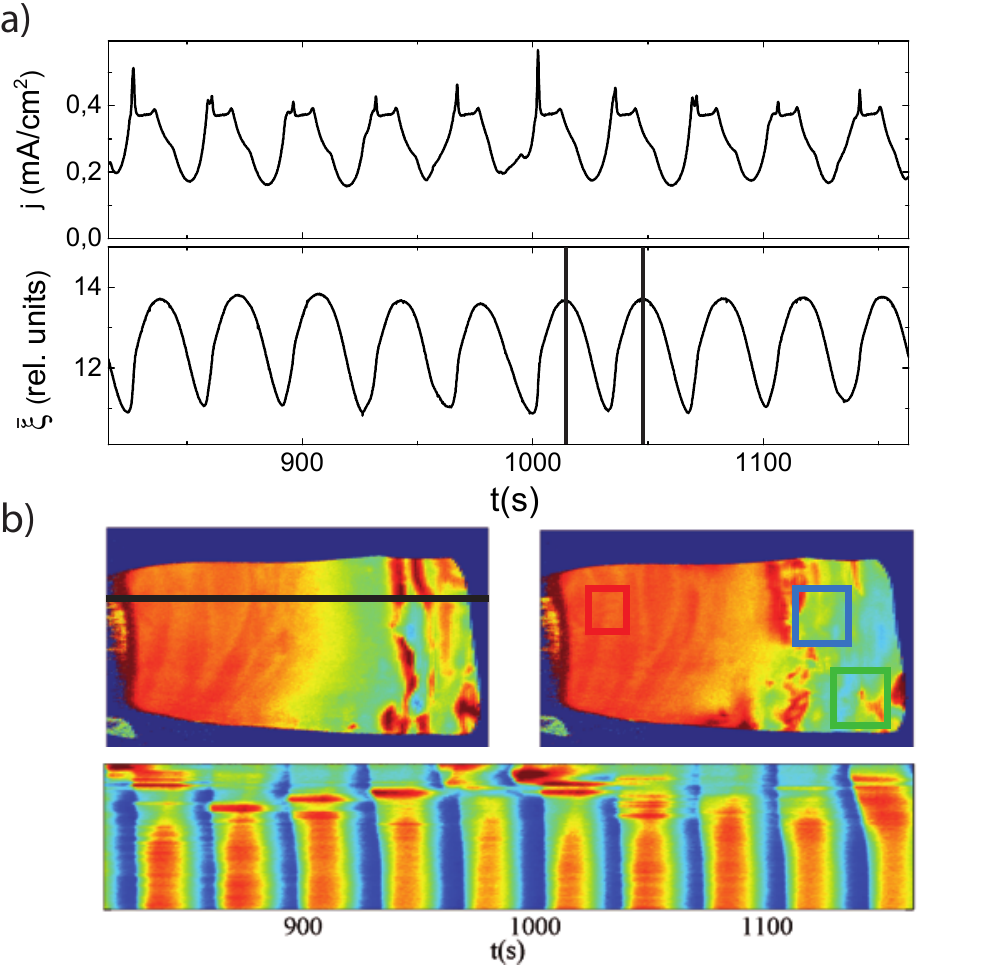}
		\caption{Analysis of a pattern consisting of a synchronized and a desynchronized region ($c_{\mathrm{F}}=75$ mM, $\mathrm{pH}=3.5$, $R_{\mathrm{ext}}A=5.61~\mathrm{k\Omega cm^2}$, $I_{\mathrm{ill}}=1.8$ mW/cm$^2$). \textbf{a)} Time series of $j$ and $\overline{\xi}$; \textbf{b)} Ellipsometric intensity distribution on the electrode for the two times indicated by the vertical lines in a) and the temporal evolution of a 1d cut along the line indicated in the left electrode picture. Red indicates a relatively high and blue a relatively low ellipsometric intensity.}
	\label{fig:Hom+Turb1}
\end{figure}

\begin{figure}[ht]
	\centering
		\includegraphics[width=0.35\textwidth]{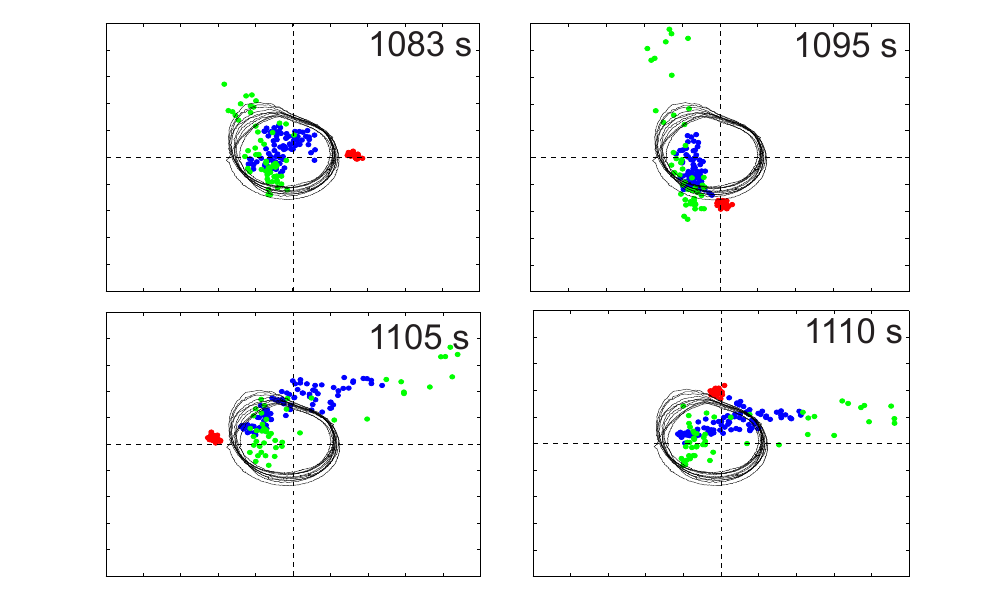}
		\caption{Analytical signal from the points indicated by correspondingly color coded rectangles in figure \ref{fig:Hom+Turb1} b) at four points during one oscillation period of the $\overline{\xi}$ oscillation shown in figure \ref{fig:Hom+Turb1} a) together with the trace of the analytical signal of $\overline{\xi}$ (solid).}
	\label{fig:Hom+Turb2}
\end{figure}

\begin{figure}[ht]
	\centering
		\includegraphics[width=0.35\textwidth]{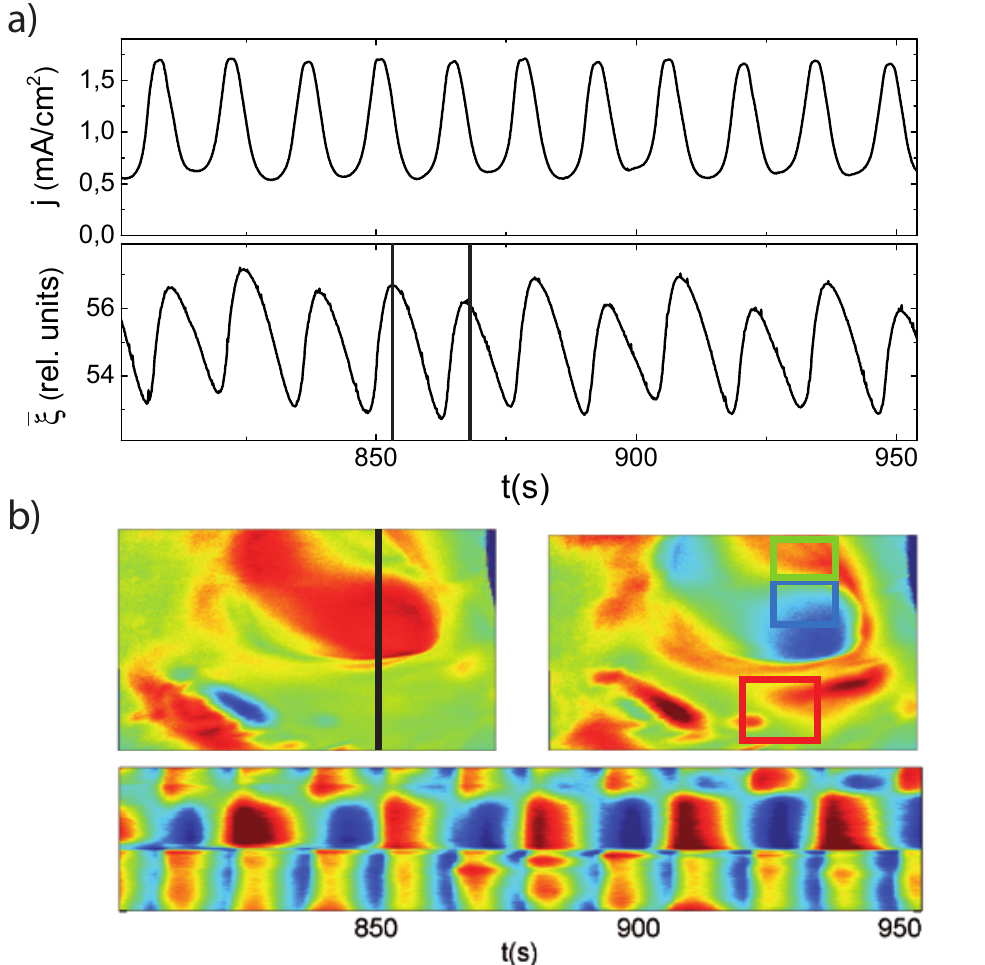}
		\caption{Analysis of a pattern consisting of a 2-phase cluster and a desynchronized region ($c_{\mathrm{F}}=100$ mM, $\mathrm{pH}=3.5$, $R_{\mathrm{ext}}A=2.81~\mathrm{k\Omega cm^2}$, $I_{\mathrm{ill}}=2.7$ mW/cm$^2$). \textbf{a)} Time series of $j$ and $\overline{\xi}$; \textbf{b)} Ellipsometric intensity distribution on the electrode for the two times indicated by the vertical lines in a) and the temporal evolution of a 1d cut along the line indicated in the left electrode picture. Red indicates a relatively high and blue a relatively low ellipsometric intensity.}
	\label{fig:SH+Turb1}
\end{figure}

\begin{figure}[ht]
	\centering
		\includegraphics[width=0.35\textwidth]{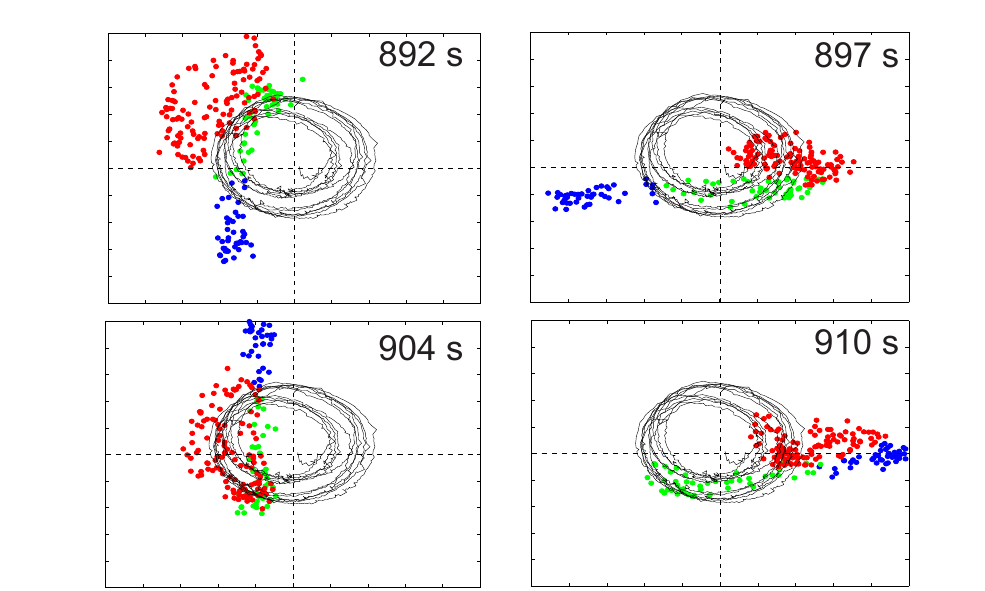}
		\caption{Analytical signal from the points indicated by correspondingly color coded rectangles in figure \ref{fig:SH+Turb1} b) at four points during two oscillation periods of the $\overline{\xi}$ oscillation shown in figure \ref{fig:SH+Turb1} a) together with the trace of the analytical signal of $\overline{\xi}$ (solid).}
	\label{fig:SH+Turb2}
\end{figure}

A region of one distinct dynamical behavior is typically not covering the entire electrode. More often two such regions are observed to coexist on the electrode. In figure \ref{fig:Hom+SH1} an example for the coexistence of a 2-phase cluster with an approximately synchronous oscillation with the frequency of the spatially averaged signal is depicted.
The two regions are clearly distinguishable in figure \ref{fig:Hom+SH1} b) with the 2-phase cluster showing a remarkably fine structured border between its two phase domains. Note that the spatially averaged signals still oscillate in a simple periodic manner.\\\\
The Fourier analysis of this pattern is shown in figure \ref{fig:Hom+SH2}.
Outside the clustered domain the subharmonic frequency is inactive as can be seen in figure \ref{fig:Hom+SH2} b). As in the experiment where the 2-phase cluster covers the entire electrode shown in figures \ref{fig:NurSH1}-\ref{fig:NurSH3}, the domain walls are Ising-type walls. The analytical signal of points from three areas on the electrode is shown in figure \ref{fig:Hom+SH3}.
Two main results can be read from the analytical signal in figure \ref{fig:Hom+SH3}. First, the center of the rotating bar formed by the two domains of the 2-phase cluster and their separating wall is now approximately stationary in the origin. This means that the underlying harmonic frequency of the spatially averaged signal is strongly suppressed in the region of the 2-phase cluster. Second, the trajectory of the points taken from the approximately synchronized region does not follow the trace of the spatially averaged analytical signal perfectly. This can also be seen in the one dimensional cut in figure \ref{fig:Hom+SH1} b) where the active phase seems to propagate across the surface. We treat this, however, as a minor variation of a perfectly uniform oscillating region. Actually, the region oscillating with the frequency of the spatially averaged signal in a synchronous manner tends to this wave-like disturbance easily.\\\\
An interesting point arises comparing the measurement where the 2-phase cluster covers the entire electrode shown in figures \ref{fig:NurSH1}-\ref{fig:NurSH3} to the measurement showing coexistence of the 2-phase cluster and synchronized, harmonic oscillation shown in figures \ref{fig:Hom+SH1}-\ref{fig:Hom+SH3}. The transition from the former to the latter dynamical state is seemingly achieved by spatially decomposing the two active frequencies. This is best seen by the corresponding deconvolution of the movement of the bar in the analytical signal shown in figure \ref{fig:NurSH3} into a pure rotation of the bar around the origin and a pure center of mass rotation around the origin in the different regions on the electrode as shown in figure \ref{fig:Hom+SH3}. The frequency ratio between the center of mass movement and the rotation of the bar remains unaffectedly 1:0.5. A further investigation of this possible deconvolution process would be desirable.\\\\ 
The next case frequently found in the experiments is the coexistence of a synchronized and a desynchronized region on the electrode. This dynamical state is typically called 'chimera state' and has attracted a lot of research interest recently \cite{Kuramoto:2002, Abrams:2004, Hagerstrom:2012, Tinsley:2012, Martens:2013, Kiss:2013}. While in most experiments in literature these states have to be initialized in a special way, their spontaneous occurrence in our system is especially interesting \cite{Smart:2012}. An exemplary 'chimera state' is shown in figure \ref{fig:Hom+Turb1}. 
Note that while a region of desynchronized local oscillations can easily be seen in figure \ref{fig:Hom+Turb1} b), the spatial average is again unaffected. Furthermore, the region of synchronized oscillations seems to be less perturbed by the presence of the region of desynchronized oscillations than by the presence of the subharmonic 2-phase cluster as discussed for figure \ref{fig:Hom+SH1}. The coexistence of the two dynamical regions is further emphasized in figure \ref{fig:Hom+Turb2} where the analytical signal of points from three areas on the surface shows the qualitative difference between the synchronized and the desynchronized oscillations.
\\\\
Another type of coexistence of synchronized and desynchronized patches on the electrode arises in the case of the coexistence of a 2-phase cluster with a desynchronized region as shown in figure \ref{fig:SH+Turb1}.
Here, the individual domains in the 2-phase cluster are again synchronized, show a phase difference of $\pi$ and are separated by Ising-type walls. This is further elucidated in the analytical signal obtained from points spreading across the Ising-type wall in figure \ref{fig:SH+Turb2} (blue and green points). Comparable to figure \ref{fig:NurSH3} these points form a bar that rotates at half the angular frequency of the $\overline{\xi}$ oscillation. In contrast, the points from the desynchronized region, displayed in red, spread out in a cloud. 
As this pattern state is again showing the coexistence of synchronized and desynchronized regions one might argue that it could also be called a 'chimera state'. It has to be noted that a slight period-2 perturbation is visible in the oscillation of $\overline{\xi}$ in figures \ref{fig:SH+Turb1} and \ref{fig:SH+Turb2}. This point will be further discussed in section \ref{subsec:P2}.\\\\ 
The coexistence of 2-phase clusters with a synchronous simple periodic oscillation as well as chimera states discussed above have also been observed in two other contexts. Tinsley et al. observed them experimentally in a network of coupled chemical oscillators and reproduced the dynamics with a reduced two group model \cite{Tinsley:2012}. Here the two groups were obtained by the introduction of a non-local feedback. Furthermore, the uniform oscillation was always stable and thus special initial conditions had to be prepared. This is in contrast to the work described here as well as to the predictions of the modified complex Ginzburg-Landau equation where the patterns occur spontaneously in a purely globally coupled system \cite{Schmidt:2014}.
     
\subsection{Period-2 oscillations in $\overline{\xi}$}
\label{subsec:P2}
\begin{figure}[ht]
	\centering
		\includegraphics[width=0.35\textwidth]{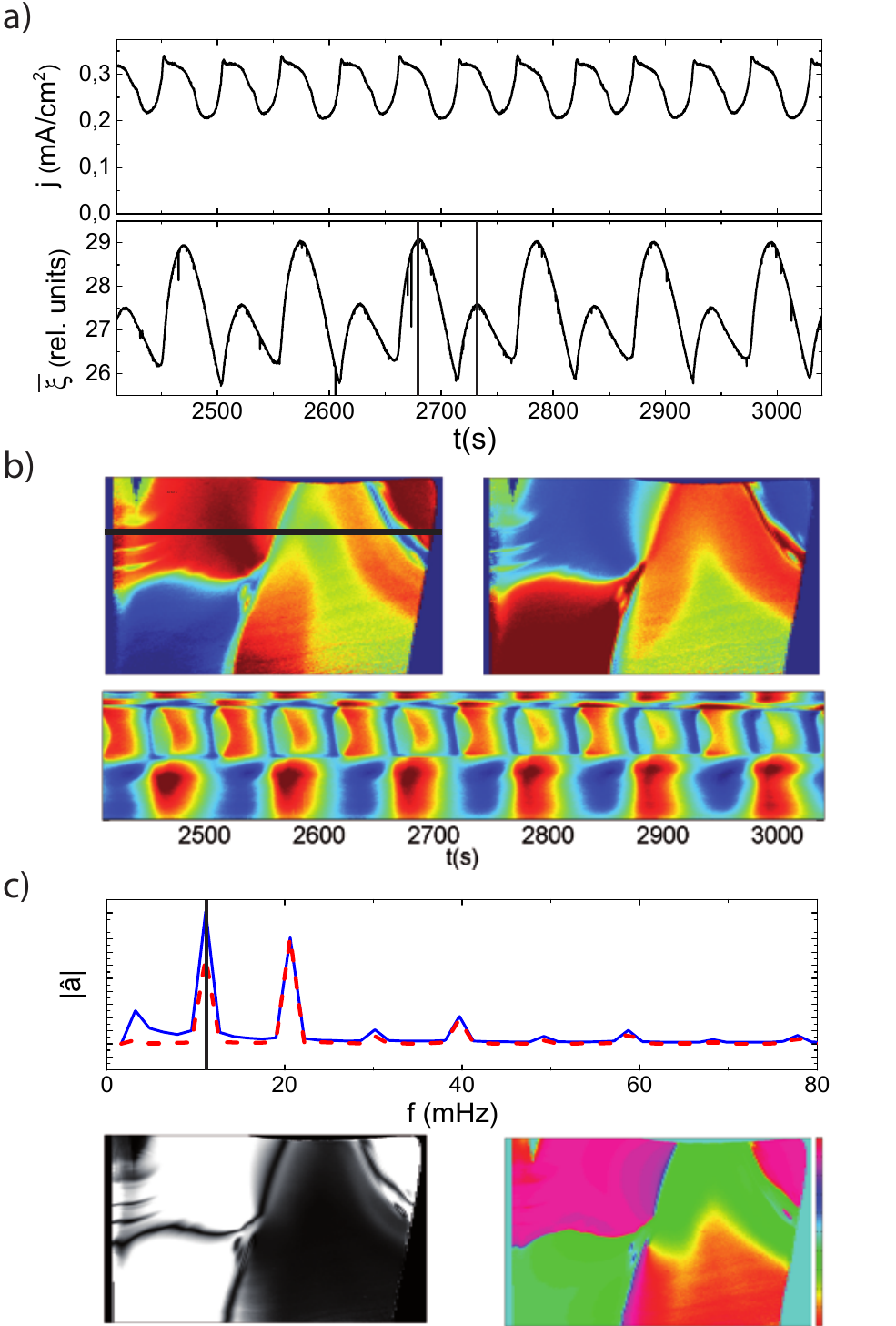}
		\caption{Analysis of a pattern consisting of a 2-phase cluster and a desynchronized region ($c_{\mathrm{F}}=50$ mM, $\mathrm{pH}=1$, $R_{\mathrm{ext}}A=2.61~\mathrm{k\Omega cm^2}$, $I_{\mathrm{ill}}=0.4$ mW/cm$^2$). \textbf{a)} Time series of $j$ and $\overline{\xi}$; \textbf{b)} Ellipsometric intensity distribution on the electrode for the two times indicated by the vertical lines in a) and the temporal evolution of a 1d cut along the line indicated in the left electrode picture. Red indicates a relatively high and blue a relatively low ellipsometric intensity; \textbf{c)} Fourier analysis of the pattern shown in a). Spatial average of the absolute values of the pointwise Fourier spectra (blue) and Fourier spectrum of the spatially averaged signal $\overline{\xi}$ (red) and the spatial distribution of the absolute value (left) and phase (right) of the Fourier coefficient corresponding to the Fourier mode indicated by the vertical line in the Fourier spectra.}
	\label{fig:SHinAv}
\end{figure}

\begin{figure}[ht]
	\centering
		\includegraphics[width=0.35\textwidth]{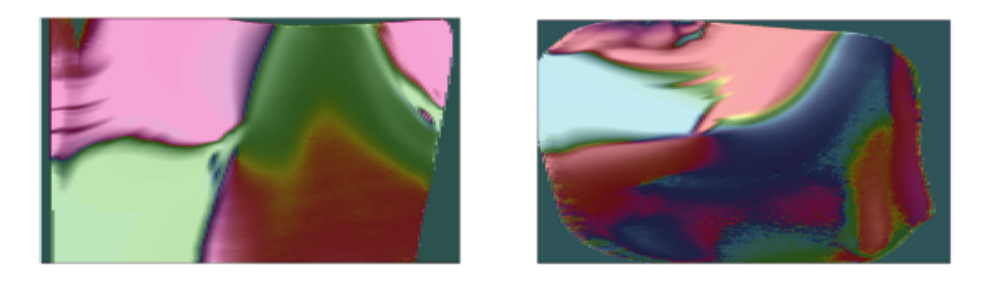}
		\caption{Comparison between the spatial distribution of absolute value and phase of the Fourier coefficients of a subharmonic cluster with different phase domain sizes and one with identical phase domain sizes.}
	\label{fig:Vergleich}
\end{figure}

A first deviation from the simple periodic oscillations in $\overline{\xi}$ is the occurrence of period-2 oscillations in $\overline{\xi}$ as shown in the middle of figure \ref{fig:Classification}. In all experiments where this behavior is found it is accompanied by the formation of a 2-phase cluster region where one phase domain covers a larger area or has a higher amplitude in the subharmonic mode than the other. An example is shown in figure \ref{fig:SHinAv}. 
In the distribution of amplitude and phase of the Fourier coefficient of the subharmonic mode in the lower images of figure \ref{fig:SHinAv} c) the lack of balance between the phase domains of the 2-phase cluster becomes apparent. The difference between the unbalanced and the balanced case is visualized in figure \ref{fig:Vergleich} by an overlay of the amplitude and phase plots of the Fourier coefficients of the subharmonic mode for the patterns shown in figures \ref{fig:SHinAv} (unbalanced) and \ref{fig:Hom+SH1} (balanced), respectively.
While the two phase domains on the left hand side of the unbalanced electrode in figure \ref{fig:Vergleich} are of similar size and the amplitude of the Fourier coefficient in these domains is similar, there is no counterbalancing phase domain for the domain in the upper right corner. For this reason the phase color coded as purple in figure \ref{fig:SHinAv} c) is also visible as an amplitude variation in the spatially averaged ellipsometric intensity signal. Whenever the average signal $\overline{\xi}$ of a patterned electrode oscillates with a period-2, unbalanced 2-phase clusters are present.

\subsection{Irregular oscillations in $\overline{\xi}$}
\label{subsec:Irreg}
\begin{figure}[t]
	\centering
		\includegraphics[width=0.35\textwidth]{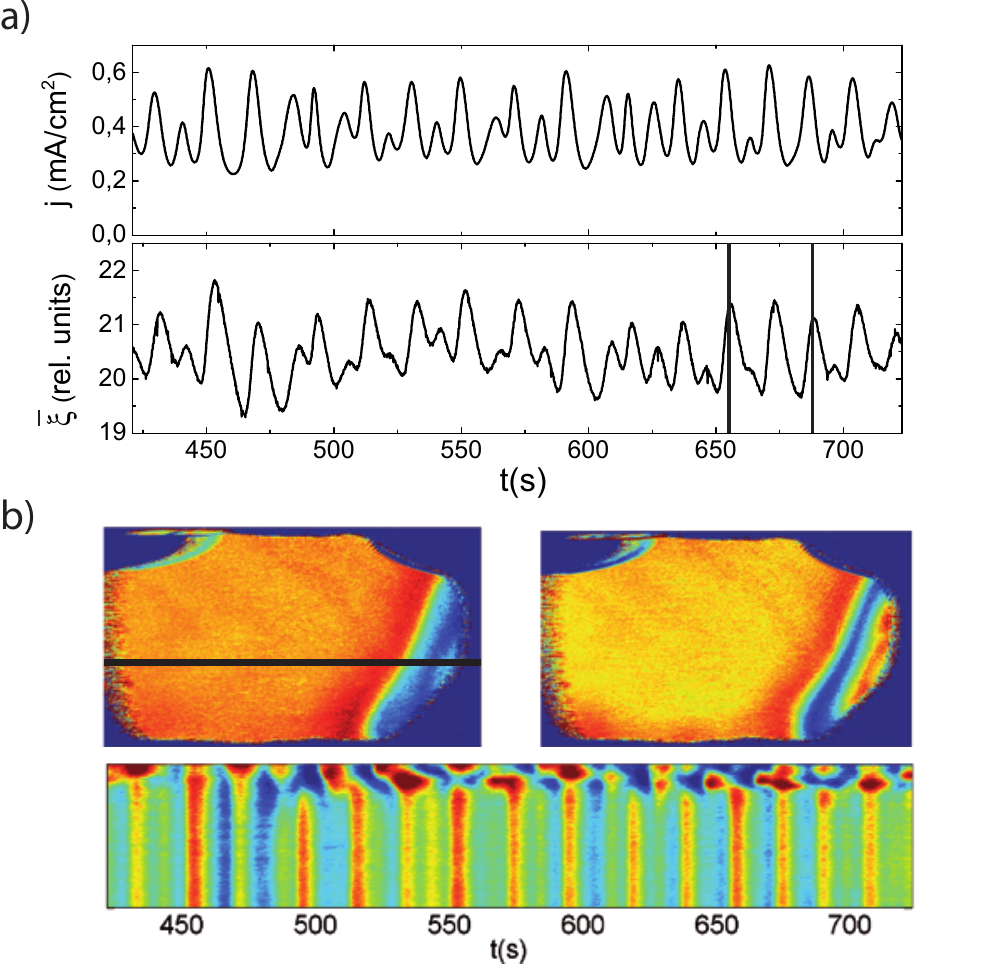}
		\caption{Analysis of a chimera type pattern occurring during an irregular $\overline{\xi}$ oscillation ($c_{\mathrm{F}}=75$ mM, $\mathrm{pH}=1$, $R_{\mathrm{ext}}A=6.51~\mathrm{k\Omega cm^2}$, $I_{\mathrm{ill}}=3.0$ mW/cm$^2$). \textbf{a)} Time series of $j$ and $\overline{\xi}$; \textbf{b)} Ellipsometric intensity distribution on the electrode for the two times indicated by the vertical lines in a) and the temporal evolution of a 1d cut along the line indicated in the left electrode picture. Red indicates a relatively high and blue a relatively low ellipsometric intensity.}
	\label{fig:Irreg1}
\end{figure}

\begin{figure}[t]
	\centering
		\includegraphics[width=0.35\textwidth]{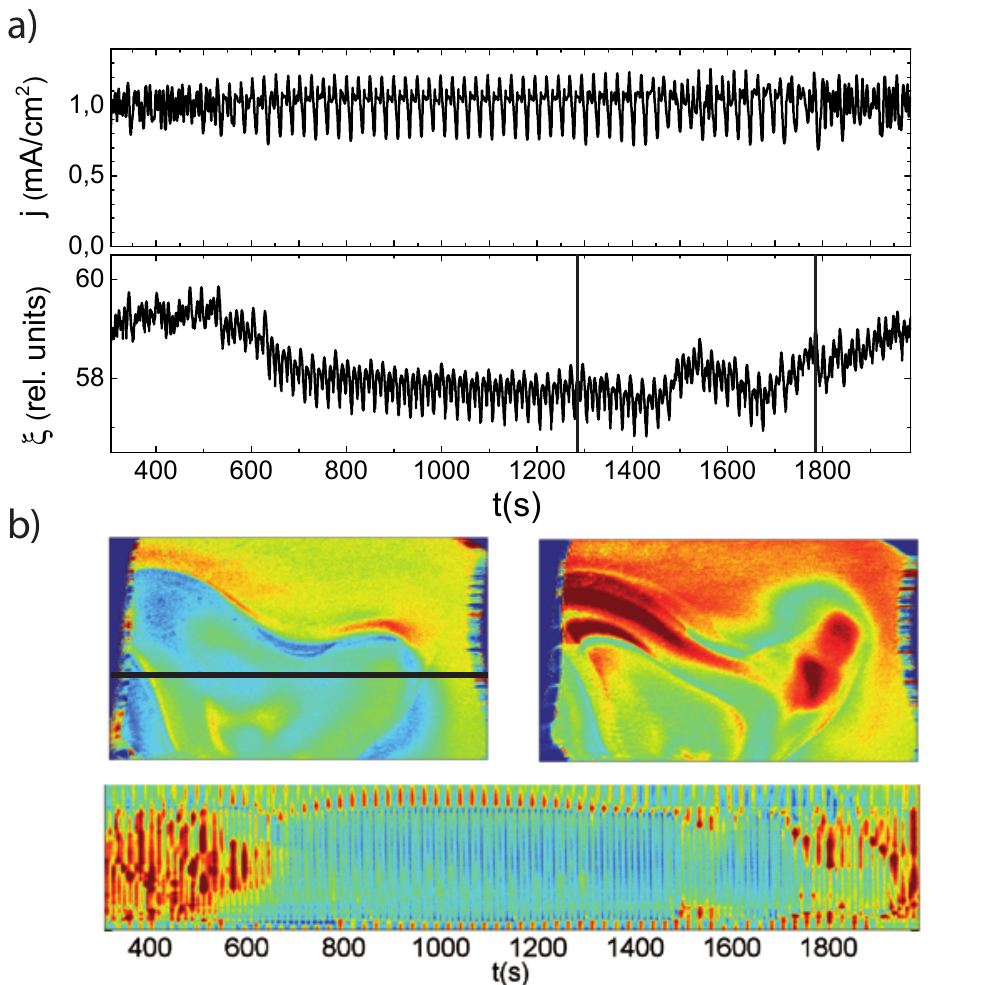}
		\caption{Analysis of a spatio-temporal pattern showing temporarily changing dynamical behavior within regions with stable walls ($c_{\mathrm{F}}=60$ mM, $\mathrm{pH}=2.3$, $R_{\mathrm{ext}}A=2.41~\mathrm{k\Omega cm^2}$, $I_{\mathrm{ill}}=1.0$ mW/cm$^2$). \textbf{a)} Time series of $j$ and $\overline{\xi}$; \textbf{b)} Ellipsometric intensity distribution on the electrode for the two times indicated by the vertical lines in a) and the temporal evolution of a 1d cut along the line indicated in the left electrode picture. Red indicates a relatively high and blue a relatively low ellipsometric intensity.}
	\label{fig:Irreg2}
\end{figure}

\begin{figure}[t]
	\centering
		\includegraphics[width=0.35\textwidth]{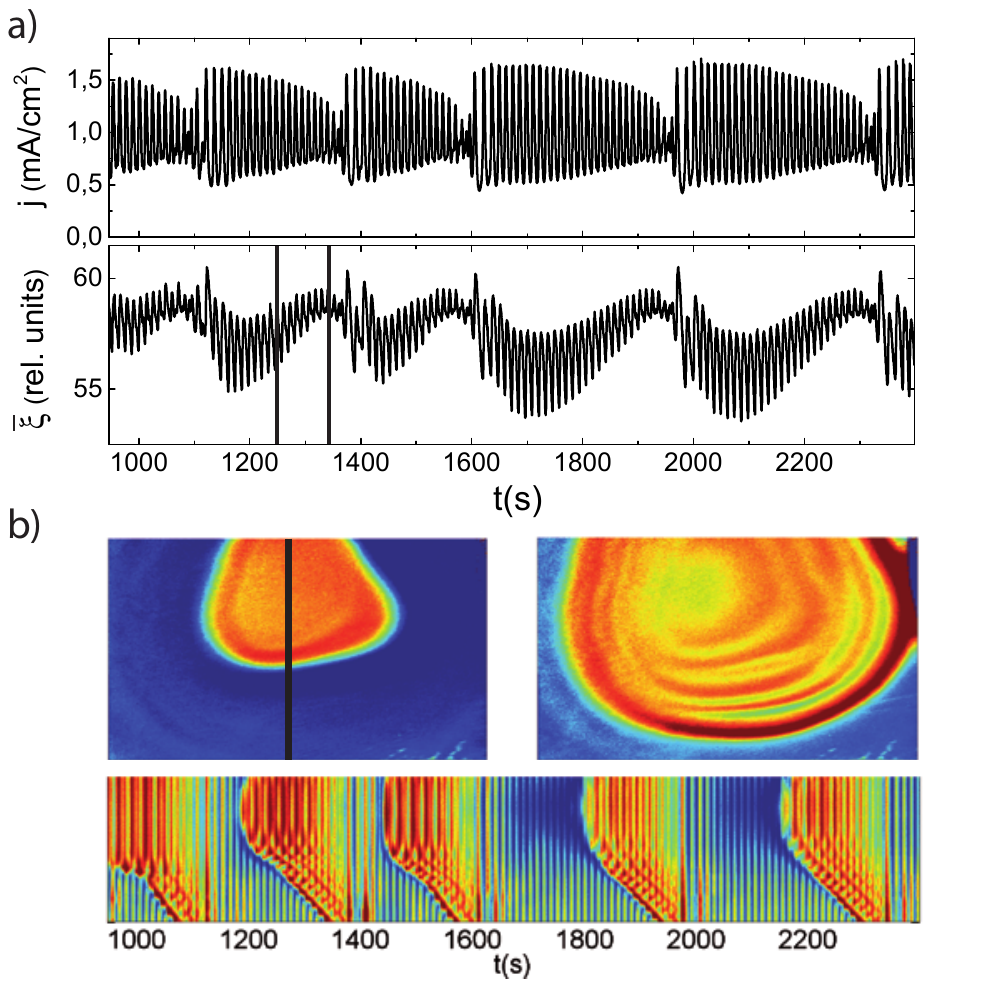}
		\caption{Analysis of a spatio-temporal pattern showing periodical changes in the region sizes ($c_{\mathrm{F}}=100$ mM, $\mathrm{pH}=3.5$, $R_{\mathrm{ext}}A=4.22~\mathrm{k\Omega cm^2}$, $I_{\mathrm{ill}}=2.7$ mW/cm$^2$). \textbf{a)} Time series of $j$ and $\overline{\xi}$; \textbf{b)} Ellipsometric intensity distribution on the electrode for the two times indicated by the vertical lines in a) and the temporal evolution of a 1d cut along the line indicated in the left electrode picture. Red indicates a relatively high and blue a relatively low ellipsometric intensity.}
	\label{fig:Irreg3}
\end{figure}

The last type of patterns discussed here occurs accompanied by an irregular oscillation of $\overline{\xi}$ as shown in the bottom of figure \ref{fig:Classification}. In this case three different pattern categories are distinguished. The first category, shown in figure \ref{fig:Irreg1}, can again be seen as a deviation from the case of simple periodic spatial average oscillations.
Here the irregular oscillation of the spatial average is imposed on the entire electrode and the relative intensity of the maxima and minima of $\overline{\xi}$ in figure \ref{fig:Irreg1} a) can be traced in the synchronized region in the lower part of the one dimensional cut in figure \ref{fig:Irreg1} b). A striking similarity between this type of pattern and the patterns forming accompanied by a simple periodic oscillation in $\overline{\xi}$ is that the regions on the electrode retain their respective dynamical behaviors and sizes until the measurement is finished. Spatially uniform oscillations with an irregular amplitude are also encountered with p-type \cite{Miethe:2012, Schoenleber:2012} and highly illuminated n-type silicon samples, i.e., in the absence of the nonlinear coupling. For this reason the measurement shown in figure \ref{fig:Irreg1} can be seen as a deviation from the 'chimera state' shown in figure \ref{fig:Hom+Turb1}. The former is caused by the nonlinear coupling imposed on a spatially uniform oscillation with irregular, most likely chaotic, dynamics, while the latter is caused by the nonlinear coupling acting on a simple periodic, spatially uniform oscillation.\\\\
The other two categories of patterns forming during oscillations of $\overline{\xi}$ with an irregular amplitude are qualitatively different from all other types of patterns considered so far. The most important difference is that the regions spontaneously forming on the electrode, while still well distinguishable, show dynamical behavior varying with time. A first example of such spatio-temporal pattern formation is shown in figure \ref{fig:Irreg2}.  
One important feature the spatio-temporal pattern shown in figure \ref{fig:Irreg2} shares with all spatio-temporal patterns shown so far, is that the regions with distinct dynamical behavior do not change their shape as is well visible in the temporal evolution of the cross section in figure \ref{fig:Irreg2} b). Conversely, the dynamical behavior in the region in the middle of the cross section changes significantly twice on a long time scale from a completely desynchronized oscillation pattern to a nearly synchronized oscillation pattern with a significantly lower amplitude and back. These profound changes in the overall behavior are also visible in the time series of $\overline{\xi}$.\\
Another qualitatively different behavior where the walls of the regions themselves are moving is shown in figure \ref{fig:Irreg3}.
In this example the respective sizes of the two regions distinguishable on the electrode surface change as one region first grows at the expense of the other to completely vanish once it fills up the entire electrode. It is remarkable in this example how long the memory of the surface seems to be as the period of the growing process is about 30 times larger than the base oscillation period.  
\section{Summary and Conclusions}
In this work, we presented a variety of self-organized spatio-temporal patterns forming in the ellipsometric intensity during the oscillatory photoelectrodissolution of n-type silicon under limited illumination. Remarkably, different regions of distinct dynamical behavior formed spontaneously under spatially uniform experimental parameters which were kept constant in time. We classified these patterns according to the behavior of the spatially averaged signal of the ellipsometric intensity $\overline{\xi}$, the dominant case being a simple periodic oscillation of $\overline{\xi}$. In this case, the patterns included synchronized oscillations, subharmonic 2-phase clusters emerging from the uniform oscillation by a period doubling bifurcation and desynchronized oscillations. Any pairwise coexistence of such dynamical regions was found including so-called 'chimera states'.
A short overview of the patterns is given in table 1. 
\begin{table}%
\centering
\begin{tiny}
\begin{tabular}{c|c|c|c|}
                    & $\overline{\xi}$ Period-1 & $\overline{\xi}$ Period-2 & $\overline{\xi}$ Irregular   \\\hline
                    & Fig.\ref{fig:NurHom1}  &                           &                              \\
Uniform             & Similar to p-Si        &       -                   &          -                   \\ 
                    &                        &                           &                              \\\hline
                    & Fig.\ref{fig:NurSH1}   &                           &                              \\
2-Ph. Cl.           & Subharm. clusters      &        -                  &      -                       \\
                    & Phase balance          &                           &                              \\\hline
                    &                        &                           &                              \\
Desync.             & Fig.\ref{fig:NurTurb1} &       -                   &      -                       \\
                    &                        &                           &                              \\\hline
                    & Fig.\ref{fig:Hom+SH1}  &  Fig.\ref{fig:SHinAv}     &    Fig.\ref{fig:Irreg3}      \\
Sync. +2-Ph.Cl.     &Phase balance           & No phase bal.             & Growing regions              \\
                    &Stat.region walls      & Stat.reg. walls          &  Mov. reg. walls             \\\hline
                    & Fig.\ref{fig:Hom+Turb1}&                           &    Fig.\ref{fig:Irreg1}      \\
Sync. +Desync.      & Chimera                &         -                 & Sync.,t-chaotic reg.         \\
                    &Stat.region walls      &                           & Stat.region walls           \\\hline
                    & Fig.\ref{fig:SH+Turb1} &                           &                              \\
2-Ph.Cl.+Desync.    & Cl.region: Ph.bal.     &         -                 &        -                     \\
                    & Stat.region walls     &                           &                              \\\hline      
\end{tabular}
\end{tiny}
\caption{Summary of the patterns and of some important properties shown in this work. The lines are sorted by the types of patterns encountered in the respective regions and the columns by the behavior of the spatially averaged signal of the ellipsometric intensity $\overline{\xi}$. (Sync.=synchonized, 2-Ph.Cl.=2-Phase Clusters, Desync.=desynchronized, Stat.=spatially stationary, reg.=region, Cl.=Cluster, Ph.=Phase)}
\label{tab:summ}
\end{table}
The conservation of the simple periodic oscillation of $\overline{\xi}$ and $j$ suggest the presence of a nonlinear global coupling. Indeed, the patterns only occur when the total coupling was found to be dominated by the nonlinear coupling imposed by the limited illumination. However, whether the coupling induced by the limited illumination of the sample is indeed global, remains an open question at this point. In addition to the case of simple periodic oscillations in $\overline{\xi}$, patterns accompanied by period-2 and irregular oscillations in $\overline{\xi}$ have been presented. In the former case, unbalanced 2-phase clusters were always present on the electrode, linking the period-2 oscillation of $\overline{\xi}$ to the lack of phase balance. The latter case opens a wide variety of possible patterns which were classified by the stability of the region walls and their long term behavior but not yet studied in detail.\\\\
The presented occurrence of patterns where different dynamical states coexist under completely uniform experimental parameters broadens the spectrum of self-organized patterns in oscillatory media. So far, mainly one particular state amongst these coexistence states, the 'chimera state', has received immense interest. Here, we demonstrate that this state is embedded in a series of related coexistence patterns, arguably showing a similar level of seemingly contradictory behavior in the coexisting regions. An example is the coexistence of a cluster and a desynchronized region shown in figure \ref{fig:SH+Turb1}. The wide variety of such patterns in the presence of a simple periodic global oscillation is remarkable. This conserved oscillation of the spatially averaged quantities seems to strongly promote the spontaneous emergence of the coexistence of different dynamical states on the electrode. In a wider sense, this can be seen as diversification of the dynamical behavior of a system with uniform parameters. The essential dynamical features responsible for this behavior remain unclear at this point. Considering our experiments, the assumption of a nonlinear global coupling as a necessary ingredient seems to be justified when considering the minority charge carrier dynamics in the silicon electrode. The lifetime of these charge carriers is several orders of magnitude shorter than any characteristic time scale of the observed oscillatory dynamics while the diffusion length is comparable to the system size within one order of magnitude. This should bring about a long-range if not effectively global coupling.\\   
In conclusion, the novel type of spatio-temporal patterns presented here sheds light on a number of fundamental questions concerning universal laws of pattern formation in oscillatory media, inviting future studies in both experiment and theory.           
\section*{Acknowledgments}
We thank our Bachelor students Martin Wiegand and Elmar Mitterreiter for carrying out some of the experiments shown and Iljana Miethe and Lennart Schmidt for fruitful discussions. Financial support from DFG (Grant KR1189/12-1) and the cluster of excellence 'Nanosystems Initiative Munich' (NIM) is gratefully acknowledged.


\end{document}